\documentclass[aps,prb,twocolumn]{revtex4}
\usepackage{graphicx}
\usepackage{dcolumn}
\usepackage{amsmath}
\usepackage{amssymb}

\def\uu{\uparrow\uparrow}
\def\ud{\uparrow\downarrow}

\def\dd{\downarrow\downarrow}
\def\u{\uparrow}
\def\d{\downarrow}
\def\rv{{\bf r}}

\def\kfr{k_{\rm F}r}
\def\kf{k_{\rm F}}

\def\fcut{F_{\rm cut}}
\def\glr{g_{\rm LR}}
\def\gosc{g_{\rm oscill}}
\def\ec{\epsilon_c}

\def\beq{\begin{equation}}
\def\eeq{\end{equation}}
\def\bear{\begin{eqnarray}}
\def\eear{\end{eqnarray}}
\begin{document}
\title{Pair-distribution functions of the two-dimensional
electron gas}
\author{Paola Gori-Giorgi,\footnote{present address: Laboratoire de Chimie Th\'eorique,
Universit\'e Pierre et Marie Curie, Paris, France} Saverio Moroni, and Giovanni B. Bachelet}
\affiliation{INFM Center for
  Statistical Mechanics and Complexity and
Dipartimento di Fisica, Universit\`a di Roma ``La Sapienza'', 
Piazzale A. Moro 2, 00185 Rome, Italy}
\date{\today}
\begin{abstract}
Based on its known exact properties and a new set of extensive fixed-node reptation
quantum Monte Carlo simulations (both with and without backflow correlations, which
in this case turn out to yield negligible improvements), we propose a new analytical
representation of (i) the spin-summed pair-distribution function and (ii) the
spin-resolved potential energy of the ideal two-dimensional interacting electron 
gas for a wide range of electron densities and spin polarization, plus (iii) the 
spin-resolved pair-distribution function of the unpolarized gas. These formulae
provide an accurate reference for quantities previously not available in analytic form,
and may be relevant to semiconductor heterostructures, metal-insulator transitions
and quantum dots both directly, in terms of phase diagram and spin susceptibility, and
indirectly, as key ingredients for the construction of new two-dimensional spin density
functionals, beyond the local approximation.
\end{abstract}
\maketitle
\section{Introduction and main results}
The two-dimensional electron gas (2DEG), realized in semiconductor
heterostructures, has been a source of lasting inspiration for at least two
generations of fundamental and applied researchers.\cite{twodgas}
In recent years, for example, a new gold rush has been triggered by the
experimental discovery of a metallic phase at low temperature,\cite{krav}
in contrast with the scaling theory of localization in two dimensions (2D),
\cite{anderson} and, independently, by the scientific and technological
progress on quantum dots, which, at semiconductor interfaces, become
nothing but tiny, quasi-two-dimensional quantum disks. \cite{dots}

In this context, accurate predictions obtained from a simplified model, such
as the ideal 2DEG (strictly 2D electrons interacting via a $1/r$ potential within
a uniform, rigid neutralizing background), represent a valuable reference.
For example, a recent analytic representation of quantum Monte Carlo correlation
energies\cite{AMGB} as a function of spin polarization $\zeta$ and coupling
parameter $r_s=1/{\sqrt{\pi n}a_B}$ (where $n$ is the density and $a_B$ is the
Bohr radius) has been immediately picked by several authors, either because
of its relevance to the phase diagram of the 2DEG,\cite{phase_diag} or because
of the corresponding prediction for the spin suceptibility,\cite{zhu,MM} or, 
last but not least, because the analytic representation of the correlation energy 
versus $n$ and $\zeta$ is a key ingredient for the density functional theory of
quantum dots.\cite{dots,excdft,LSD2Dvarie}\par
Such an interest encouraged us to extend our previous work on energies
to the spin-resolved pair-distribution functions $g_{\sigma \sigma'}({\bf r},{\bf r}')$ 
of the 2DEG, whose accuracy and availability in analytic form may serve a variety
of purposes: the exchange-correlation hole and its dependence on the electron
density and spin polarization may be relevant to the physics of the 
metal-insulator bifurcation in 2D\cite{GN} and to self-energy theories of the
2DEG,\cite{VZNeeds} but is also needed for the estimate of the effects
of the finite thickness on the
spin susceptibility\cite{stefania} and for the construction of generalized-gradient
approximations (GGA) or weighted-density approximations (WDA) of density
functionals, in analogy to the 3D case.\cite{GGA,WDA}
The availability of density functionals better than local-spin-density (LSD)
approximations for the 2DEG would, in turn, allow an almost exact description
of quantum dots, since the spatial variation of their carrier density is rather
weak.\cite{dots,LSD2Dvarie}\par

In this paper, we exploit the known exact properties of the pair-distribution
functions (recalled in Sec.~\ref{defin_exact}), and, based on a new set of
extensive fixed-node quantum Monte Carlo simulations (described
in Sec.~\ref{sec_QMC}), we propose, in Sec.~\ref{sec_fittone}, our analytic
representation of (A) the spin-summed pair-distribution function of the ideal
two-dimensional interacting electron gas for a wide range of electron densities
and spin polarization, and (B) the spin-resolved pair-distribution function of
the unpolarized gas. In Sec.~\ref{risultati} we discuss the quality of such an
interpolation, and finally, in Sec.~\ref{energiapot} , we evaluate
the spin-resolved potential energy, of interest in the construction of
dynamical exchange-correlation potentials in the spin
channel,\cite{vignale,quian}
and propose the corresponding analytic representation.\par

As a result, quantities which are relevant to the physics of semiconductor 
heterostructures and quantum dots, and/or represent a key ingredient for the 
construction of two-dimensional spin density functionals beyond the local
approximation, are now available, for the first time, in analytic form.

Fortran subroutines for the evaluation of the parametrized quantities are available
upon request to gp.giorgi@caspur.it or Giovanni.Bachelet@roma1.infn.it.
\section{Definitions and Exact Properties}
\label{defin_exact}
For an electronic system, the pair-distribution functions $g_{\sigma
\sigma'}({\bf r},{\bf r}')$, if $n_{\sigma}({\bf r})$ is the
density of electrons with spin $\sigma=\uparrow$ or $\downarrow$, are
defined as
\beq
g_{\sigma \sigma'}({\bf r},{\bf r}')= \frac{
 \langle\Phi|\psi_{\sigma}^{\dagger}({\bf r})
 \psi_{\sigma'}^{\dagger}({\bf r}')
\psi_{\sigma'}({\bf r}')\psi_{\sigma}({\bf r})|\Phi\rangle}
{n_{\sigma}({\bf r})n_{\sigma'}({\bf r}')},
\label{g_def}
\eeq
where $\psi_{\sigma}^{\dagger}$ and $\psi_{\sigma}$ are the 
creation and annihilation field operators, respectively, and
$\Phi$ is the ground-state wavefunction.
The functions $g_{\sigma\sigma'}$ are thus related to the probability 
of finding two electrons of prescribed spin orientations at positions 
${\bf r}$ and ${\bf r}'$. The normalization is such that the
case of completely independent particles
(without exchange and correlation) corresponds to
the condition $g_{\sigma\sigma'}=1$. Hartree atomic units are used
throughout this work. 

For a two-dimensional uniform electron gas, the functions $g_{\sigma\sigma'}$
only depend on $r=|\rv-\rv'|$, and parametrically on the density
parameter $r_s=1/\sqrt{\pi\,n}$ and on the spin-polarization parameter
$\zeta=(n_\u-n_\d)/n$. The total (spin-summed) pair-distribution function
is defined as
\beq
g=\left(\tfrac{1+\zeta}{2}\right)^2g_{\uu}+
\left(\tfrac{1-\zeta}{2}\right)^2g_{\dd}+
\tfrac{1-\zeta^2}{2}\,g_{\ud}.
\eeq

For small $r$, when two electrons get closer and closer, the behavior
of $g_{\sigma\sigma'}$ is governed by the cusp conditions,\cite{Kimball2} 
\begin{eqnarray}
\frac{\partial}{\partial r}g_{\uparrow \downarrow}(r,r_s,\zeta)\biggr|_{r = 0}
& = &  2\,g_{\uparrow \downarrow}(r= 0,r_s,\zeta) 
\label{cusp_ud} \\
\frac{\partial}{\partial r}g_{\sigma\sigma}(r,r_s,\zeta)\biggr|_{r = 0}
& = & g_{\sigma\sigma}(r= 0,r_s,\zeta) = 0
\label{Pauli} \\
\frac{\partial^3}{\partial r^3}g_{\sigma\sigma}(r,r_s,\zeta)\biggr|_{r = 0}
& = & 2\,
\frac{\partial^2}{\partial r^2}g_{\sigma\sigma}(r,r_s,\zeta)\biggr|_{r = 0}.
\label{cusp_uu}
\end{eqnarray}
Eqs.~(\ref{cusp_ud}) and (\ref{cusp_uu}) are due to the dominance
of the potential term $1/|\rv-\rv'|$ in the many-body hamiltonian
as $\rv \to \rv'$; Eq.~(\ref{Pauli}) comes from the Pauli principle. 

At this point, it is convenient to introduce the scaled variable $x=\kfr$, where
 $\kf=\sqrt{2}/r_s$ is the Fermi wavevector of the unpolarized gas.

The Fourier transforms of $g_{\sigma\sigma'}-1$ are the 
spin-resolved static structure factors,\cite{pines} which, for a 2D uniform gas,
are 
\beq
S_{\sigma\sigma'}(q,r_s,\zeta)=\delta_{\sigma\sigma'}+
\frac{\sqrt{n_\sigma n_{\sigma'}}}{n}\int_0^{\infty} dx 
\,[g_{\sigma\sigma'}-1]\,x\,J_0(qx),
\eeq  
where $q=k/\kf$ is a scaled variable in reciprocal space, and
$J_0$ is the Bessel function of order 0. The total
(spin-summed) static structure factor is 
\beq
S=\tfrac{1+\zeta}{2}\,S_{\uu}+\tfrac{1-\zeta}{2}\,S_{\dd}
+\sqrt{1-\zeta^2}\,S_{\ud};
\eeq
its long-wavelength (i.e., small-$q$) behavior is 
determined by the plasma collective mode,\cite{pines}
\beq
S(q\to 0,r_s,\zeta)=\frac{q^{3/2}}{2^{3/4}r_s^{1/2}}+O(q^2),
\label{eq_plasmon}
\eeq
and thus does not depend on $\zeta$.

Usually $g_{\sigma\sigma'}$ (and consequently $S_{\sigma\sigma'}$) is 
conventionally divided into the (known) exchange and the
(unknown) correlation terms,
\bear
 & & g_{\sigma\sigma'}  =  g_{\sigma\sigma'}^x+g_{\sigma\sigma'}^c, \\
 & & g_{\ud}^x  =   1, \\
& &  g_{\sigma\sigma}^x  =  1-\left[\frac{2\,J_1(\kf^{\sigma} r)}{\kf^\sigma r}
\right]^2, \\
& & S_{\sigma\sigma'}  =  S_{\sigma\sigma'}^x+S_{\sigma\sigma'}^c, \\
& &  S_{\ud}^x  =  0, \\
& & S_{\sigma\sigma}^x  = \frac{2}{\pi}\left[\arcsin\left
(\frac{k}{2\kf^{\sigma}}\right)+
\frac{k}{2\kf^\sigma}
\sqrt{1-\left(\frac{k}{2\kf^\sigma}\right)^2}\right] 
\times \nonumber \\
 & & \qquad \qquad \times \theta(2\kf^\sigma-k)+\theta(k-2\kf^\sigma), 
\label{eq_Sx}
\eear
where $J_1$ is the first-order Bessel function, $\theta$ is the Heaviside
step function, and $\kf^\u=\kf\sqrt{1+\zeta}$, $\kf^\d=\kf\sqrt{1-\zeta}$.
The functions $g^x$ and $S^x$ correspond to a uniform 2-dimensional system of
noninteracting fermions; once the scaled variables $x$ and $q$ are used,
they do not depend explicitly on $r_s$:
$g^x=g^x(x,\zeta)$, $S^x=S^x(q,\zeta)$. In what follows, we use the name
{\em pair-distribution} function for the whole thing ($g=g^x+g^c$, exchange 
plus correlation), and {\em pair-correlation} function for its correlation-only 
contribution $g^c$.

Combining Eqs.~(\ref{eq_plasmon}) and (\ref{eq_Sx}), we find the small-$q$
behavior of the spin-summed correlation static structure factor,
\beq
S^c(q\to 0,r_s,\zeta) = -\frac{2}{\pi}\phi(\zeta)\,q+
\frac{q^{3/2}}{2^{3/4}r_s^{1/2}}+O(q^2),
\label{eq_Scsmallq}
\eeq
where
\beq
\phi(\zeta)=\frac{\sqrt{1+\zeta}+\sqrt{1-\zeta}}{2}
\label{eq_phi}
\eeq
plays the same role of the three-dimensional function $\phi$ of
Refs.~\onlinecite{PW} and \onlinecite{GP2}.
As well known from the properties of Fourier transforms,
the small-$q$ behavior of $S$ determines the 
oscillation-averaged long-range part of $g$.
We thus see that, individually taken, $g^c$ and $g^x-1$ have long-range
tails $\propto r^{-3}$; but these tails exactly cancel in the pair-distribution
function (exchange plus correlation), so that $g-1=g^x+g^c-1$ approches
zero as $r^{-7/2}$. 

While the long-wavelength limit of the total $S$, Eq.~(\ref{eq_plasmon}),
is well known, little is known about the small-$q$ behavior of the 
spin-resolved $S_{\sigma\sigma'}$ (and hence
about the long-range part of $g_{\sigma\sigma'}$). 
The conservation of the number of  particles implies
\beq
S_{\sigma\sigma'}(q=0,r_s,\zeta)=0.
\label{eq_srss}
\eeq
In Section~\ref{sec_fit} we discuss an approximate expression
for $S_{\sigma\sigma'}(q\to 0,r_s,\zeta=0)$ consistent with our QMC results.

Finally, the spin-summed $g^c$ yields the correlation part of the 
expectation value of the Coulomb potential energy, $v_c(r_s,\zeta)$,
which can be obtained from the correlation energy $\ec(r_s,\zeta)$
via the virial theorem,\cite{virial}
\beq
\frac{\kf}{2}\int_0^{\infty} dx\,g^c(x,r_s,\zeta) = v_c(r_s,\zeta)=
\frac{1}{r_s}\frac{\partial}{\partial r_s}\left[
r_s^2\ec(r_s,\zeta)\right].
\label{eq_vc}
\eeq

\section{Quantum Monte Carlo calculation}
\label{sec_QMC}
The ground-state expectation value ${\bar {\cal O}}$ of a local 
operator ${\hat {\cal O}}$, such as the pair-distribution function or the 
static structure factor, is estimated as
\beq
{\bar {\cal O}}=\langle\Psi(\beta)|{\hat{\cal O}}|\Psi(\beta)\rangle
/\langle\Psi(\beta)|\Psi(\beta)\rangle
\label{pure-estimators}
\eeq
using a reptation quantum Monte Carlo (RQMC) algorithm.\cite{rqmc}
Here $\Psi$ is a trial function, and 
$\Psi(\beta)={\rm e}^{-\beta H/2}\Psi$ can be made sufficiently close to
the exact ground state $\Phi$ by choosing the ``imaginary time'' 
$\beta$ large enough.

The estimate of Eq.~(\ref{pure-estimators}) is called ``pure'', as opposed
to the ``mixed'' estimate ${\bar {\cal O}}_{mix}=
\langle\Phi|{\hat{\cal O}}|\Psi\rangle /\langle\Phi|\Psi\rangle$
usually adopted in connection with the diffusion Monte Carlo (DMC)
method.\cite{mitas} 
More precisely, previous DMC results for the
pair-distribution function of the 2D electron gas\cite{tanatar,rapisarda,kwon}
have been based on extrapolated estimates,
${\bar {\cal O}}_{ext}=2{\bar {\cal O}}_{mix}
-\langle\Psi|{\hat{\cal O}}|\Psi\rangle /\langle\Psi|\Psi\rangle$. 
The bias in ${\bar {\cal O}}_{ext}$ is quadratic in the error of the 
trial function. Such an estimate is often very accurate, but
a well converged pure estimate, as obtained in the present work,
has the advantage of being independent of the quality of the 
trial function $\Psi$ (except for its nodal structure, see below).

The RQMC method features a discretized path integral representation of 
the importance-sampled imaginary time propagator, 
\bear
{\tilde G}(R_0&\to& R_P;\beta)=\Psi(R_P)
\langle R_P|{\rm e}^{-\beta H}|R_0\rangle/\Psi(R_0)\\
&=&\int dR_1\cdots dR_{P-1} \Pi_{i=0}^{P-1}
{\tilde G}(R_i\to R_{i+1};\epsilon),\nonumber
\label{path-integral}
\eear
where $\epsilon=\beta/P$ is the time step and $R_i$ is the set of
the $2N$ coordinates of the $N$ electrons at the $i$-th step.
We use the standard short-time approximation\cite{mitas}
\bear
{\tilde G}(R\to R';\epsilon)&\simeq& A 
{\rm e}^{-[R'-R-\epsilon\nabla\ln\Psi(R)]^2/2\epsilon}\\
&\times&{\rm e}^{-\epsilon[E_L(R')+E_L(R)]/2},\nonumber
\label{drift-diffusion}
\eear
where $E_L(R)=H\Psi(R)/\Psi(R)$ is the ``local energy'', and
$A=(2\pi\epsilon)^{-N}$ is a normalization constant.
Replacement of Eqs.~(\ref{path-integral}) and (\ref{drift-diffusion})
into (\ref{pure-estimators}) yields an integral amenable to
Monte Carlo evaluation, using a generalized Metropolis algorithm to
sample paths in an enlarged configuration space, $X=\{R_0,\cdots,R_P\}$. 

In our simulations we consider $N_{\uparrow}$ spin-up and 
$N_{\downarrow}$ spin-down particles in a square box with periodic 
boundary conditions. The spin-resolved pair-distribution functions are
obtained\cite{ortiz} averaging 
$V dN_{\sigma\sigma'}(r')/[N_\sigma(N_{\sigma'}
-\delta_{\sigma\sigma'})2\pi r\Delta]$
in the middle slice of the path during the simulation, 
where $V$ is the volume of the simulation cell, $dN_{\sigma\sigma'}(r')$
is the number of electron pairs with distance $r'$ between $r-\Delta/2$
and $r+\Delta/2$. The structure factors are computed analogously, 
for vectors ${\bf k}$ in the reciprocal lattice of the simulation cell, by
averaging 
$\rho_\sigma({\bf k})\rho_{\sigma'}(-{\bf k})/(N_\sigma N_{\sigma'})^{1/2}$,
where $\rho_\sigma({\bf k})=\sum_j \exp(-i{\bf k}\cdot{\bf r}_j)$ is 
the density fluctuation of electrons with spin $\sigma$.
The total number of particles is 42, 50, 50 and 45 for polarization
0, 0.48, 0.80 and 1, respectively. By repeating simulations for different
system sizes in the unpolarized case, finite size effects on the 
pair-distribution function have been estimated to be of order 0.01.
The systematic bias due to finite projection
time and finite time step can be kept within this
level by suitable choices of the parameters $\beta$ and $\epsilon$.
In our simulations, this results in paths of 501 time slices.

We avoid the fermion sign problem using the fixed node 
approximation (FNA),\cite{mitas} whereby the paths
are not allowed to cross the nodes of the trial function.
The FNA, which gives the lowest energy upper bound
consistent with the nodal structure of the trial function, is
the only source of uncontrolled approximation in the present
calculation. In order to gauge the sensitivity of the computed
pair-distribution function on the nodal structure of $\Psi$, we
have performed our simulations using two trial functions with 
different nodes. 

Our first trial function is of the simplest Jastrow-Slater form
$\Psi(R)\!=\!J(R) S(R)$. Here $J(R)\!=\!\Pi_{i<j}\exp(-u(r_{ij}))$, $r_{ij}$ 
being the distance between the electrons $i$ and $j$, is a symmetric
Jastrow factor; it describes pair correlations through the function
$u(r)$, which is optimized (by minimizing the variational energy)
for each density and polarization; it's always positive, so it does 
not alter the nodal structure, which is entirely determined by
the other factor $S(R)$, a product of two Slater determinants
(one for each spin component) of plane-wave one-particle 
orbitals $\exp({\bf k}_i\cdot{\bf r}_j)$.

Our second trial function has the same Jastrow factor, but
its nodal structure is more accurate, since it includes 
``backflow'' correlations\cite{kwon2,kwon} by replacing the 
electron coordinates ${\bf r}_j$ in the Slater determinants 
with ``quasi-coordinates''
\beq
{\bf x}_j={\bf r}_j+\sum_{i\neq j}\eta(r_{ij})({\bf r}_i-{\bf r}_j),
\label{backflow}
\eeq
where $\eta(r)$ is another function to be optimized for each
density and polarization.

\begin{figure}
\includegraphics[width=7.cm]{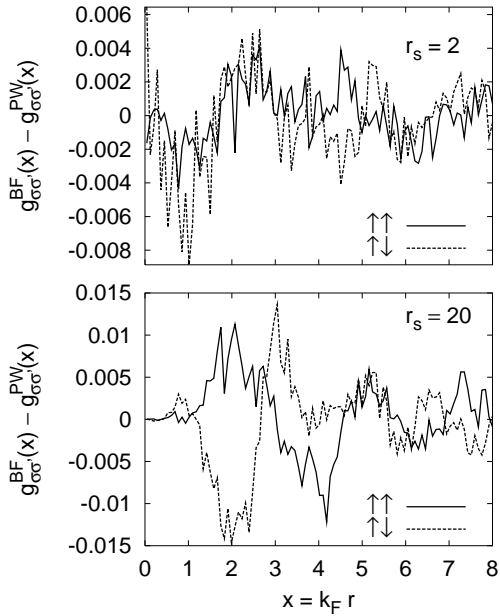}
\caption{Numerical difference between the spin-resolved pair-distribution 
functions $g_{\sigma\sigma'}^{BF}-g_{\sigma\sigma'}^{PW}$ at $r_s\!=\!2$
(upper panel) and $r_s\!=\!20$ (lower panel), as obtained from two fixed-node
simulations with different nodal structures. The superscript indicates
backflow ($BF$) or plane-wave ($PW$) nodes.}
\label{b_diff}
\end{figure}

\begin{figure}
\includegraphics[width=8.5cm]{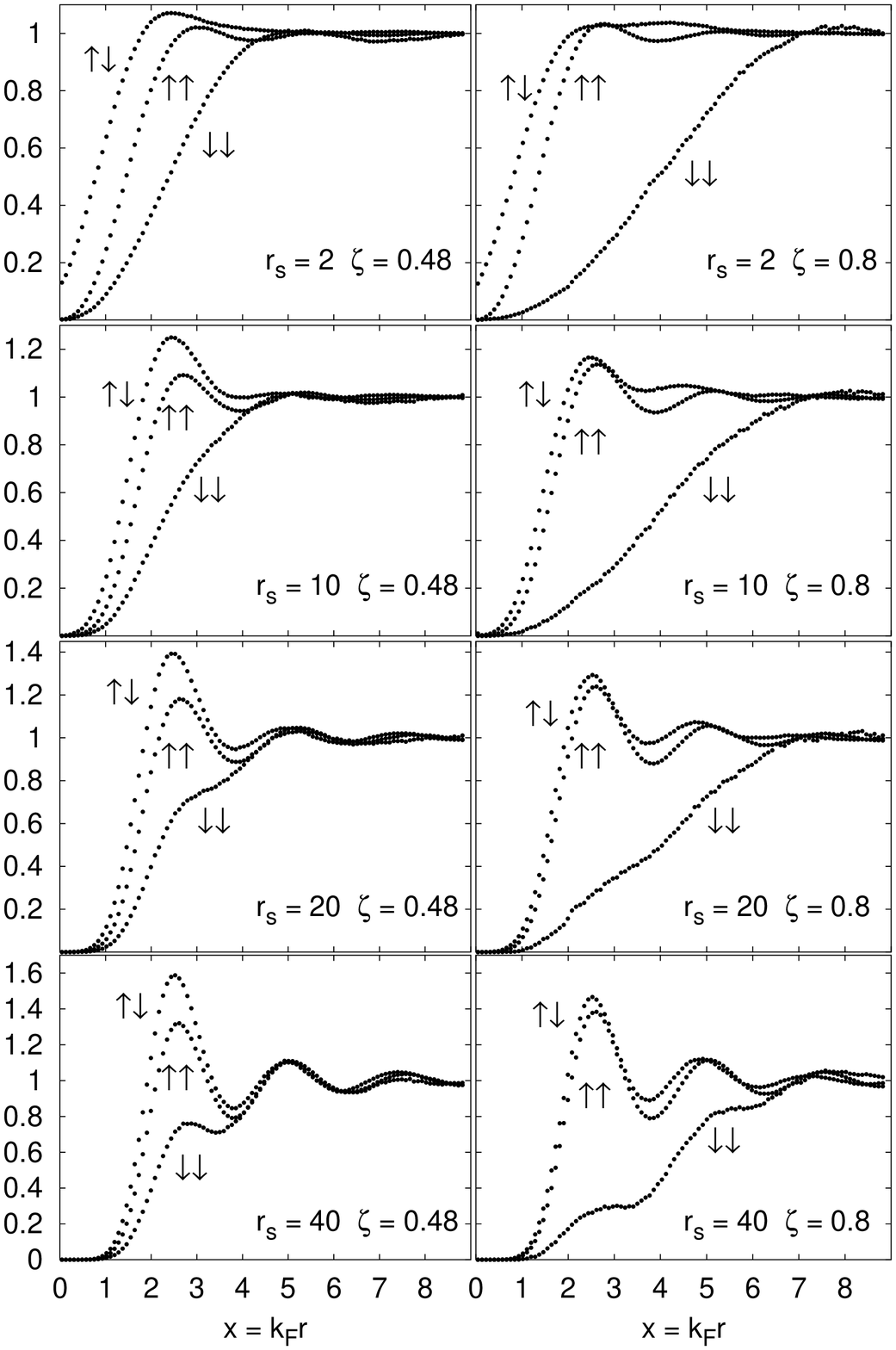} 
\caption{Sample of spin-resolved pair-distribution functions as directly
obtained from our QMC simulations (no fitting here).}
\label{fig_qmc}
\end{figure}

In a previous variational calculation\cite{kwon} the difference
between the pair-distribution function calculated with the simple
Slater-Jastrow and the backflow trial function was found to be of
order 0.03. Here we find that, in a fixed-node calculation, such effect
is even smaller:
Figure~\ref{b_diff} shows the difference in $g_{\uparrow\uparrow}$
and $g_{\uparrow\downarrow}$, computed with either plane-wave or 
backflow nodes, for $r_s=2$ and 20 at zero polarization. In the
worst case (large $r_s$, lower panel) these differences are half
as large as found in the variational case,\cite{kwon} while for
small $r_s$ (upper panel) they are much smaller than that.
These differences are essentially invisible on the scale  of our 
$g_{\sigma\sigma'} (r,r_s,\zeta)$ calculated with plane-wave nodes,
some samples of which are shown Figure~\ref{fig_qmc}. As a 
consequence, an analytic representation of the spin-summed $g(r,r_s,\zeta)$ 
and of $g_{\sigma\sigma'}(r,r_s,\zeta=0)$ (see next Sec.~\ref{sec_fittone}) 
based on the plane-wave results, as the one presented here, happens
to give an equally good representation of the backflow results,
because the difference due to the improved nodal structure is either 
comparable or smaller than the fitting error.

\section{Analytic representation}
\label{sec_fittone}
In this section we describe our analytic representations of
the spin-summed pair-correlation function $g^c(x,r_s,\zeta)$
valid for $1\le r_s\le 40$ and $0\le \zeta\le 1$, and
of the spin-resolved $g^c_{\sigma\sigma'}(x,r_s,\zeta)$
for $\zeta=0$ and $1\le r_s\le 10$.
These functions are built along the lines
of Refs.~\onlinecite{PW, GP2} and of Ref.~\onlinecite{GSB} for the 3D case.

The strategy is the following. We build the spin-summed
$g^c$ as a sum of three terms: long-range, short-range, and
oscillatory. The long-range term is taken from the random-phase
approximation (RPA) and multiplied by a cutoff function which quenches
its short-range contribution. The short-range part  is built
according to the cusp conditions of 
Eqs.~(\ref{cusp_ud})-(\ref{cusp_uu}), as a weighted sum of
$\uu$, $\ud$ and $\dd$ terms which, in turn,  have been determined
for $\zeta=0$ by a fitting procedure to the QMC results. For
$\zeta\neq 0$, an exchange-like $\zeta$-dependence of these
$\sigma\sigma'$ short-range coefficients has been assumed.
The oscillatory part is empirical, being
entirely determined by a fit to the QMC data. The analytic function
$g^c$ is also constrained, via Eq.~(\ref{eq_vc}), to reproduce 
our parametrized correlation energy of Ref.~\onlinecite{AMGB}.
\begin{figure}
\includegraphics[width=6.8cm]{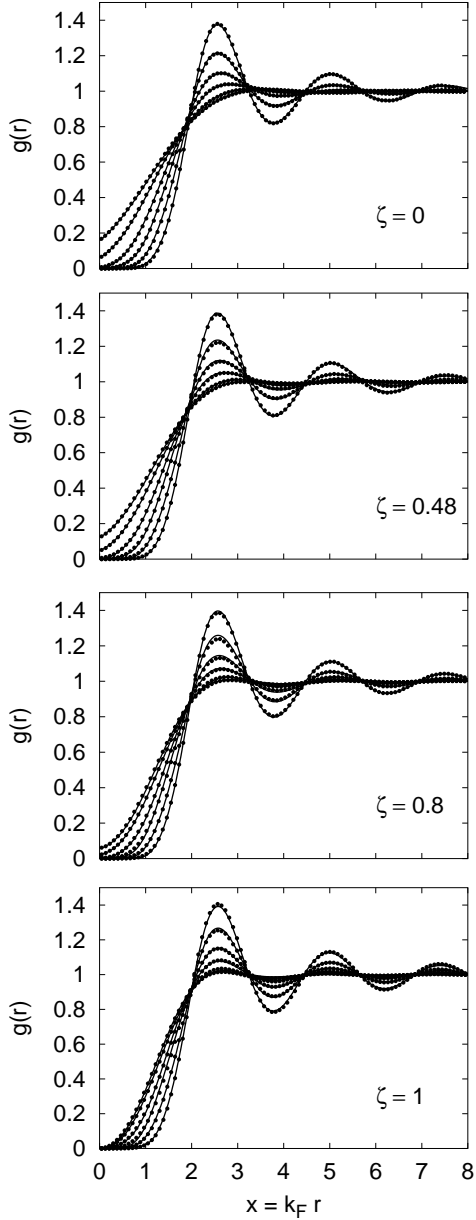} 
\caption{Spin-summed pair-distribution function (exchange plus 
correlation: $g=g^x+g^c$, see text) for four different values of the spin-polarization 
parameter $\zeta$, and for $r_s= 1$, 2, 5, 10, 20, and 40 (larger
$r_s$ values have stronger oscillations). The dots correspond to our QMC
data, the solid lines to our analytic representation. Error bars are comparable
with the dot size.}
\label{fig_totfit}
\end{figure}

The analytic parametrization of the spin-resolved $g_{\sigma\sigma'}^c$
is more difficult, because less is known about its exact properties.
We had to rely more heavily on our QMC data, and, for the time being,
we successfully interpolated $g_{\ud}^c$ only in the unpolarized
case ($\zeta\!=\!0$) and for $r_s\in [1,10]$. This parametrization, 
combined with the one
for the total $g^c$ also yields $g^c_{\uu}=g^c_{\dd}=2g^c-g^c_{\ud}$.
We build $g_{\ud}^c$ using a functional form similar to the one just
described  for the total $g^c$: a sum of a long-range term, 
a short-range term, and an oscillatory term. The long-range term
is obtained by a modification, consistent with our QMC data, of the 
long-range analytic form appropriate for the total (spin-summed) $g^c$.
The short-range term is simply the $\ud$ part of the total $g^c$. The
oscillatory part is, again, empirical. Because the short-range
parts of the total $g^c$ and of $g_{\ud}^c$ share some parameters,
we performed a simultaneous, global, three-dimensional ($x,r_s,\zeta$) fit of
$g^c(x,r_s,\zeta)$ and  $g^c_{\ud}(x,r_s,\zeta=0)$.
This procedure and all the relevant equations are detailed in the next
subsections.    
\subsection{Spin-summed pair-correlation function}
We parametrize the spin-summed $g^c$ as
\beq
g^c=[g_{\rm LR}(x)+g_{\rm oscill}(x)]\,\fcut(x)
+e^{-d\,x^2}\sum_{n=0}^6
c_n x^n,
\label{eq_gtot}
\eeq
where $\glr$ is a long-ranged function whose Fourier transform exactly
recovers Eq.~(\ref{eq_Scsmallq}), $\gosc$ is an oscillating
function to be fitted to the QMC data, 
and the last term in the r.h.s takes care of 
the short-range properties. The function $F_{\rm cut}(x)$ 
quenches\cite{GP2}
the short-range contribution of $(g_{\rm LR}+g_{\rm oscill})$,
\beq
\fcut = 1-e^{-d\,x^2}\left(1+d\,x^2+\tfrac{1}{2}d^2x^4+
\tfrac{1}{6}d^3x^6\right) \, .
\label{eq_fcut}
\eeq
The parameter $d(r_s)$ determines
the mixing of long-range and short-range terms in
Eq.~(\ref{eq_gtot}).
\subsubsection{Long-Range part}
The long-range part is built with the same procedure used
for the 3D case in Refs.~\onlinecite{PW,GP2} and~\onlinecite{WP}, 
and detailed in Appendix~\ref{app_LR},
\beq
\glr(x,r_s,\zeta)=2\,\phi^5(\zeta)\,r_s^2 \,\frac{f_1(v)}{x},
\label{eq_glr}
\eeq
where $v=\sqrt{2} r_s \phi^2 x$ is another scaled variable, and
$\phi(\zeta)$ is given by Eq.~(\ref{eq_phi}). The function
$f_1(v)$ is reported in Appendix~\ref{app_LR}.
\subsubsection{Short-Range part}
The short-range part of our $g^c$ is the last term in the r.h.s of
Eq.~(\ref{eq_gtot}). We have
\bear
 c_0 &  = &  \tfrac{1-\zeta^2}{2}\,g^c_{\ud}(0)
\label{eq_c0} \\
 c_1 &  = &  \tfrac{2}{\kf}\,\tfrac{1-\zeta^2}{2}\,[g^c_{\ud}(0)+1] \\
 c_2 &  = &  d\,c_0+\tfrac{1-\zeta^2}{2}\,a_2^{\ud}+
\left(\tfrac{1+\zeta}{2}\right)^2a_2^{\uu}+
\left(\tfrac{1-\zeta}{2}\right)^2a_2^{\dd} \nonumber \\
& & -\tfrac{1}{8}(1+3\,\zeta^2) \\
c_3 & = & d\,c_1+\tfrac{1-\zeta^2}{2}\,a_3^{\ud}+
\left(\tfrac{1+\zeta}{2}\right)^2a_3^{\uu}+
\left(\tfrac{1-\zeta}{2}\right)^2a_3^{\dd},
\label{eq_c3}
\eear
where $a_n^{\sigma\sigma'}$ are the short-range coefficients of
the spin-resolved pair-distribution functions,
\beq
g_{\sigma\sigma'}(x\to 0,r_s,\zeta)=\sum_{n} a_n^{\sigma\sigma'}
x^n,
\eeq
and
\beq
a_3^{\sigma\sigma}=\frac{2}{3\kf}a_2^{\sigma\sigma}.
\eeq
The pair-correlation function at zero electron-electron distance, or 
``on-top" value $g^c_{\ud}(0)\equiv a_0^{\ud}-1$, has been
parametrized as
\beq
g^c_{\ud}(0)=[1+(a-1.372)\,r_s+b\,r_s^2+c\,r_s^3]\,e^{-a\,r_s}-1.
\label{eq_g0}
\eeq
The parameters $a=1.46$, $b=0.258$, $c=0.00037$ are fitted
to the QMC results; the exact high-density slope, 1.372,
is taken from Ref.~\onlinecite{polini}.

As said at the beginning of this section, we determine the spin-resolved
short-range coefficients for the $\zeta=0$ case, and then we assume
an exchange-like $\zeta$-dependence. This means that in 
Eqs.~(\ref{eq_c0})-(\ref{eq_c3}) the values of $g^c_{\ud}(0)$, 
$a_2^{\ud}$ and $a_3^{\ud}$ only depend on $r_s$ (not
on $\zeta$), and that the coefficients 
$a_2^{\uu}$ and $a_2^{\dd}$ have the simple
$\zeta$ dependence 
\beq
a_2^{\uu}(r_s,\zeta)=\tfrac{1}{4}(1+\zeta)
a_p(r_s),
\label{eq_a2p}
\eeq 
with $a_2^{\dd}(r_s,\zeta)=a_2^{\uu}(r_s,-\zeta)$.

The linear parameters $c_4$ and $c_5$ will be used to constrain 
$g^c$ to yield the correlation energy of Ref.~\onlinecite{AMGB} and
to fulfill the particle-conservation sum rule [$S(q=0,r_s,\zeta)=0$], as
in Ref.~\onlinecite{GP2}. The parameter $c_6(r_s,\zeta)$ is used
to give more variational freedom to our $g^c$ for an accurate fit of  
the QMC data at higher $r_s$.
\subsubsection{Oscillatory part} 
The oscillatory part of our $g^c$ is similar to the form
used by Tanatar and Ceperley,\cite{tanatar}
\beq
\gosc=\frac{m_1}{x+1}\,e^{-m_2\,x} \cos(m_3\,x+m_4)
\label{eq_goscill}
\eeq
which is able to accurately fit the QMC data at low
densities. The exponential cutoff ensures that $\gosc$ does not
alter the long-range properties embedded in $\glr$. The parameters
$m_i$ depend on both $r_s$ and $\zeta$.
\subsubsection{Sum rules}
As said, the role of the parameters $c_4$ and $c_5$ which appear in the
short-range part of our $g_c$ is to fulfill the normalization sum rule
[$S^c(q=0)=0$] and to recover the correlation energy $\ec(r_s,\zeta)$ of
Ref.~\onlinecite{AMGB}. We obtain
\bear
c_4 & = & 8\,d^2\frac{15\sqrt{d}\sqrt{\pi}\,C_e-16\,d\,C_s}{45\pi-128} \\
c_5 & = & 16\,d^3\frac{3\sqrt{d}\sqrt{\pi}\,C_s-8\,C_e}{45\pi-128},
\eear
with
\bear
C_s & = & -\frac{c_0}{2d}-\frac{c_1\sqrt{\pi}}{4d^{3/2}}-\frac{c_2}{2d^2}
-c_3\frac{3\sqrt{\pi}}{8d^{5/2}} -\frac{3\,c_6}{d^4}\nonumber \\
& & +2\phi^5r_s^2\,s_{\rm LR}
-s_{\rm oscill} \\
C_e & = & -\frac{c_0\sqrt{\pi}}{2\sqrt{d}}-\frac{c_1}{2d}-
\frac{c_2\sqrt{\pi}}{4d^{3/2}}-\frac{c_3}{2d^2}
-c_6\frac{15}{16}\frac{\sqrt{\pi}}{d^{7/2}} \nonumber \\
& & -2\phi^5r_s^2 E_{\rm LR}
-E_{\rm oscill}+\sqrt{2}{r_s}\,v_c \\
s_{\rm LR} & = & \int_0^{\infty} f_1(v) \,[1-\fcut(x)]\,dx 
\label{eq_slr}\\
E_{\rm LR} & = & 
\int_0^{\infty} \frac{f_1(v)}{x} \fcut(x)\,dx \\
s_{\rm oscill} & = & \int_0^{\infty}\gosc(x)\,x\,\fcut(x)\,dx \\
E_{\rm oscill} & = & \int_0^{\infty}\gosc(x)\fcut(x)\,dx,
\label{eq_eoscill}
\eear
and $v_c(r_s,\zeta)$ given in Eq.~(\ref{eq_vc}). 
Equations~(\ref{eq_slr})-(\ref{eq_eoscill}) are evaluated numerically for
given $r_s$ and $\zeta$.
\begin{figure}
\includegraphics[width=6.8cm]{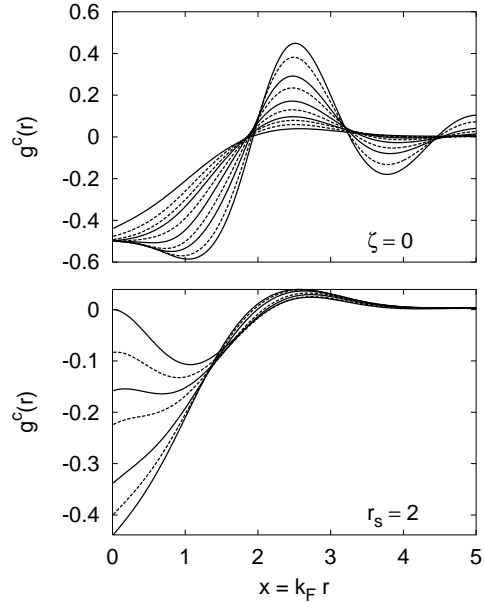} 
\caption{Spin-summed 
pair-correlation functions $g^c$ from our analytic representation. 
Upper panel: for $\zeta=0$, we show $g^c$
for $r_s=2,3,4,5,7,10,15,20,30,40$; stronger oscillations
correspond to higher $r_s$ values. The solid lines correspond
to $r_s=2,5,10,20,40$, for which our $g^c$ accurately fits the
QMC data; the dashed lines are the results for intermediate 
values of $r_s$.
Lower panel: for $r_s=2$, we show $g^c$ for different values of
the spin-polarization $\zeta=0,0.3,0.48,0.7,0.8,0.9,1$;
more negative ``on-top'' values $g^c(x=0)$ correspond to lower
values of $\zeta$. The solid
lines correspond to $\zeta=0,0.48,0.8,1$, for which our $g^c$ accurately
fits the QMC data. The dashed lines correspond to intermediate values
of $\zeta$. }
\label{fig_interp}
\end{figure}
\begin{table}
\begin{tabular}{llll}
\hline
total $g^c$: & & & \\
\hline\hline
 $\delta_1 = 0.293$ &   $\delta_2 = 0.136$  &            &   \\  
 $\gamma^{(2)}_1=0.0586$ & $\gamma^{(2)}_2=0.153$ 
& $\gamma^{(2)}_3=0.476$    & \\  
 $\gamma^{(3)}_1=0.0457$ & $\gamma^{(3)}_2=0.0427$ 
   & $\gamma^{(3)}_3=0.229$             & \\ 
 $\lambda_1=0.0377$ & $\lambda_2=0.123$ 
& $\lambda_3=0.68$    & \\  
$\beta_1=0.828$ & $\eta_1=0.11$ & $\beta_2=445$  & $\eta_2=-82$ \\
$p_1^{(1)}=3.69$ & $q_1^{(1)}=-0.987$ &  $p_2^{(1)}=4.74$ 
& $q_2^{(1)}=2.83$  \\
$p_1^{(2)}=0.92$ & $q_1^{(2)}=-0.443$ &  $p_2^{(2)}=0.044$ & 
$q_2^{(2)}=-0.0151$  \\
$p_1^{(3)}=2.14$ & $q_1^{(3)}=0.394$ &  $p_2^{(3)}=0.045$ 
& $q_2^{(3)}=-0.0299$  \\
$p_1^{(4)}=6.39$ & $q_1^{(4)}=-0.592$ &  $p_2^{(4)}=2.7\cdot10^{-4}$ 
& $q_2^{(4)}=-1.8\cdot 10^{-4}$  \\
\hline
\hline
$g^c_{\ud}$: & & & \\
\hline  
\hline
$\gamma^{(5)}_1=1.1$ & $\gamma^{(5)}_2=29$ & & \\
$\nu^{(1)}_1=0.479$ & $\nu^{(1)}_2=0.029$ & & \\ 
$\nu^{(2)}_1=0.6$ &  & & \\ 
$\nu^{(3)}_1=1.99$ & $\nu^{(3)}_2=0.0014$ & & \\ 
$\nu^{(4)}_1=1.437$ & $\nu^{(4)}_2=0.1$ & & \\ 
\hline
\hline
\end{tabular}
\caption{Optimal parameters for the analytic representation
of $g^c(x,r_s,\zeta)$ and $g^c_{\ud}(x,r_s,\zeta=0)$ as described
in Sec.~\ref{sec_fittone}.}
\label{tab_partot}
\end{table}
\subsubsection{Fitting parameters}
The parameters $d(r_s)$, $a_2^{\ud}(r_s)$, $a_3^{\ud}(r_s)$, 
$a_p(r_s)$,
$c_6(r_s,\zeta)$, $m_i(r_s,\zeta)$ are used to fit the QMC data. Their
$r_s$ and $\zeta$
dependence is smooth and allows for an analytic representation
of $g^c(x,r_s,\zeta)$ valid at all $r_s \in [1,40]$ and
$\zeta \in [0,1]$:
\bear
d(r_s) & = & \frac{\delta_1 + \delta_2 r_s^2}{1+\delta_2 r_s^2} 
\label{eq_d}\\
a_2^{\ud}(r_s) & = & (-\gamma^{(2)}_1 r_s+\gamma^{(2)}_2 r_s^2)
\,e^{-\gamma^{(2)}_3 r_s} \label{eq_a2ud}\\
a_3^{\ud}(r_s) & = & (-\gamma^{(3)}_1 r_s+\gamma^{(3)}_2 r_s^2)
\,e^{-\gamma^{(3)}_3 r_s} \label{eq_a3ud}\\
a_p(r_s) & = & (1-\lambda_1 r_s
+\lambda_2 r_s^2)\,e^{-\lambda_3 r_s} \\
c_6(r_s,\zeta) & = & \gamma^{(6)}_1(\zeta) e^{-\gamma^{(6)}_2(\zeta)/r_s^2}
\label{eq_c6} \\
m_1(r_s,\zeta) & = & \mu_1^{(1)}(\zeta) e^{-\mu_2^{(1)}(\zeta)/r_s}
\label{eq_m1} \\
m_2(r_s,\zeta) & = &  \frac{\mu_1^{(2)}(\zeta)}{1+\mu_2^{(2)}(\zeta)r_s} \\
m_3(r_s,\zeta) & = &  \frac{\mu_1^{(3)}(\zeta)+2.7\mu_2^{(3)}(\zeta)r_s}{1+\mu_2^{(3)}(\zeta)r_s} 
\label{eq_m3} \\
m_4(r_s,\zeta) & = & \frac{\mu_1^{(4)}(\zeta)+5.36\mu_2^{(4)}(\zeta)r_s^2}{1+\mu_2^{(4)}(\zeta)r_s^2}.
\label{eq_m4}
\eear
The functional form of the short-range coefficients
$a_n^{\sigma\sigma'}$ is very similar to the one used
for the 3D case in Ref.~\onlinecite{GP1}; the corresponding
parameters are determined by simultaneously fitting the data
for $g_{\ud}^c$ (see next Section) and those
for the total $g^c$. The parameter $c_6$ only comes into play
at high $r_s$: its functional form, Eq.~(\ref{eq_c6}),
makes it vanish very rapidly as $r_s$ decreases. The same argument applies 
to the 
oscillatory part, whose magnitude is determined by the parameter $m_1$ of
Eq.~(\ref{eq_m1}). The low-density limit of the parameters $m_3$ and
$m_4$, 2.7 and 5.36 in Eqs.~(\ref{eq_m3}) and~(\ref{eq_m4}), are taken from
an oversimplified model of localization on the sites of a triangular
lattice.\cite{nota} The $\zeta$ dependence of the parameters $\gamma^{(6)}_i$
and $\mu_i^{(n)}$ is well represented by a quadratic form:
\bear
\gamma^{(6)}_i(\zeta) & = & \beta_i+\eta_i \,\zeta^2 \\
\mu_i^{(n)}(\zeta) & = & p_i^{(n)}+q_i^{(n)}\,\zeta^2.
\eear
The final 32 free parameters (plus 9 parameters for
$g^c_{\ud}$, detailed in the next section) are fitted to our
data set  (100 values of $x$ for each $r_s =1,2,5,10,20,40$ 
and $\zeta=0,0.48,0.8,1$ plus those for $g^c_{\ud}$ at
$\zeta =0$ and $r_s = 1,2,5,10$ -- a total of 2800 data), and
are reported in Table~\ref{tab_partot}.
\begin{figure}
\includegraphics[width=6.8cm]{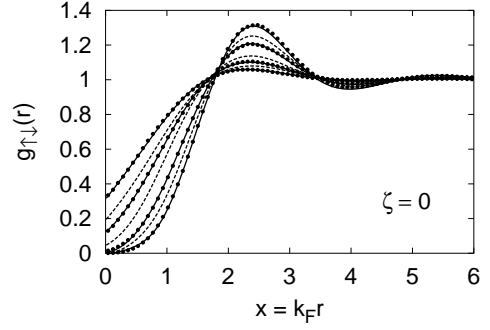} 
\caption{$\ud$ pair-distribution function 
(exchange plus correlation, see text) for $\zeta=0$
and $r_s= 1$, 1.5, 2, 3, 5, 7, 10 (the larger
$r_s$ values have stronger oscillations).
The dots correspond to our QMC data for $r_s=1$, 2, 5, 10; the solid lines is
our analytic representation at the same $r_s$-values. Dashed lines
correspond to our analytic representation for the other values of $r_s$.}
\label{fig_uduu}
\end{figure}
\subsection{Spin-resolved pair-correlation functions ($\zeta=0$)}
\label{sec_fit}
We parametrize the $\ud$ correlation function with a functional
form similar to the one used for the spin-summed $g^c$,
\beq
g^c_{\ud}=[\glr^{\ud}(x)+
g_{\rm oscill}^{\ud}(x)]\,\fcut(x)+e^{-d\,x^2}\sum_{n=0}^5
c_n^{\ud} x^n,
\label{eq_gss}
\eeq
where the function $\fcut(x)$ and the parameter $d(r_s)$ are 
given in Eqs.~(\ref{eq_fcut}) and~(\ref{eq_d}), respectively.

\subsubsection{Long-Range part}
While the long-range part of the spin-summed $g^c$, Eq.~(\ref{eq_glr}),
can be obtained from RPA, the spin-resolution is more problematic.
Nonetheless, RPA can give some hints,\cite{GSB} 
especially in the $r_s\to 0$ limit. From RPA we obtain,
up to $O(q^2)$,
\beq
S^{c\,({\rm RPA})}_{\sigma\sigma'}(q\to 0,r_s,\zeta)=
-\frac{q}{\pi}\xi_{\sigma\sigma'}(\zeta)+\frac{q^{3/2}}{2^{3/4}
r_s^{1/2}} \frac{\sqrt{n_\sigma n_{\sigma'}}}{n},
\eeq
with $\xi_{\ud}(\zeta)=1$, $\xi_{\uu}(\zeta)=2/\sqrt{1+\zeta}-
\sqrt{1-\zeta}/\sqrt{1+\zeta}$, and $\xi_{\dd}(\zeta)=\xi_{\uu}(-\zeta)$.
\par
Here, we only treat the $\zeta=0$ case, for which we also produced
spin-resolved static structure factors with QMC. We write the small-$q$
part of $S^c_{\sigma\sigma'}$ as the RPA result plus an 
$r_s$-dependent correction, similar to the 3D case,\cite{BGS}
i.e., up to $O(q^2)$,
\beq
S_{\sigma\sigma'}^c(q\to 0, r_s,\zeta=0) =
-q\left[\frac{1}{\pi}+\alpha_{\sigma\sigma'}(r_s)\right]
+\frac{q^{3/2}}{2^{7/4}r_s^{1/2}},
\label{eq_Scssz0}
\eeq
with $\alpha_{\uu}(r_s)=-\alpha_{\ud}(r_s)$. This small-$q$
behavior embodies the following properties: (i) the corresponding
spin-resolved pair-distribution function $g_{\sigma\sigma'}(r)$
are more long-ranged\cite{Atwal1} than the spin-summed $g(r)$,
(ii) parallel- and antiparallel-spin correlations give identical 
contributions to the plasma collective mode. The correction $\alpha_{\ud}(r_s)$
has been determined from the QMC results in reciprocal space
for $1\le r_s\le 10$, and is well represented by
\beq
\alpha_{\ud}(r_s)=0.00914\,r_s.
\eeq
Thus,
for the spin-resolved long-range part we use a scaling law similar to the one
of Eq.~(\ref{eq_glr}),
\beq
\glr^{\ud}(x,r_s,\zeta)=2\,\phi^5(\zeta)\,r_s^2 \,
\frac{\tilde{f}_1(v,\alpha_{\ud})}{x};
\label{eq_gsslr}
\eeq
the function $\tilde{f}_1(v,\alpha)$ is described 
in Appendix~\ref{app_LRss}.
\subsubsection{Short-Range part}
The short-range part of $g^c_{\ud}$ is the $\ud$ part of the total
$g^c$ [see Eqs.~(\ref{eq_c0})-(\ref{eq_c3})]. We thus have
\bear
c_0^{\ud} & = & g^c_{\ud}(0) \\
c_1^{\ud} & = & \frac{4}{\kf}[g^c_{\ud}(0)+1] \\
c_2^{\ud} & = & d\,c_0^{\ud}+a_2^{\ud} \\
c_3^{\ud} & = & d\,c_1^{\ud}+a_3^{\ud},
\eear
where $g^c_{\ud}(0)$, $a_2^{\ud}$, and $a_3^{\ud}$ are given in Eqs.~(\ref{eq_g0}),
(\ref{eq_a2ud}), and (\ref{eq_a3ud}), respectively.\par
The linear parameter $c_4^{\ud}$ is used to fulfill the normalization
sum rule of Eq.~(\ref{eq_srss}); the parameter $c_5^{\ud}(r_s)$, instead, increases
the variational flexibility of $g^c_{\ud}$, and is fitted to the QMC data. 
\subsubsection{Oscillatory part}
For $g^c_{\ud}$ we use the same form [Eq.~\ref{eq_goscill}] of the total $g^c$,
\beq
\gosc^{\ud}=\frac{m_1^{\ud}}{x+1}\,e^{-m_2^{\ud}\,x} 
\cos(m_3^{\ud}\,x+m_4^{\ud}).
\eeq
The parameters $m_i^{\ud}$ depend on $r_s$ and are fitted to the QMC data.
\subsubsection{Sum rule}
The sum rule Eq.~(\ref{eq_srss}) determines the linear
parameter $c_4^{\ud}$,
\vskip -0.3cm
\beq
c_4^{\ud}=d^3\,C_s^{\ud},
\eeq
\vskip -0.3cm
with
\bear
C_s^{\ud} & = & -\frac{c_0^{\ud}}{2d}-\frac{c_1^{\ud}\sqrt{\pi}}{4d^{3/2}}
-\frac{c_2^{\ud}}{2d^2}
-c_3^{\ud}\frac{3\sqrt{\pi}}{8d^{5/2}} -c_5^{\ud}\frac{15\sqrt{\pi}}{16\,d^{7/2}}
 \nonumber \\
& & +2\phi^5r_s^2\,s_{\rm LR}^{\ud}-s_{\rm oscill}^{\ud}, \vphantom{\biggl[} \\
s_{\rm LR}^{\ud} & = & \int_0^{\infty} \tilde{f}_1(v,\alpha_{\ud}) \,[1-\fcut(x)]\,dx \\
s_{\rm oscill}^{\ud} & = & \int_0^{\infty}\gosc^{\ud}(x)\,x\,\fcut(x)\,dx.
\eear
\subsubsection{Fitting parameters}
From the global fit described for the total $g^c$, we also find 
the $r_s$ dependence of the coefficients $c_5^{\ud}$ and $m_i^{\ud}$:
\bear
c_5^{\ud}(r_s) & = & \gamma^{(5)}_1 e^{-\gamma^{(5)}_2/r_s^2} \\
m_1^{\ud}(r_s) & = & \frac{\nu_1^{(1)} r_s}{1+\nu_2^{(1)} r_s} \\
m_2^{\ud}(r_s) & = & \nu_1^{(2)}  \\
m_3^{\ud}(r_s) & = & \nu_1^{(3)} +\frac{\nu_2^{(3)} r_s^2}{1+\nu_2^{(3)} r_s^2} \\
m_4^{\ud}(r_s) & = & \frac{\nu_1^{(4)}}{1+\nu_2^{(4)} r_s}. 
\eear
The values of $\gamma^{(5)}_i$ and $\nu_i^{(n)}$ are reported in
Table~\ref{tab_partot}.
\begin{figure}
\includegraphics[width=6.8cm]{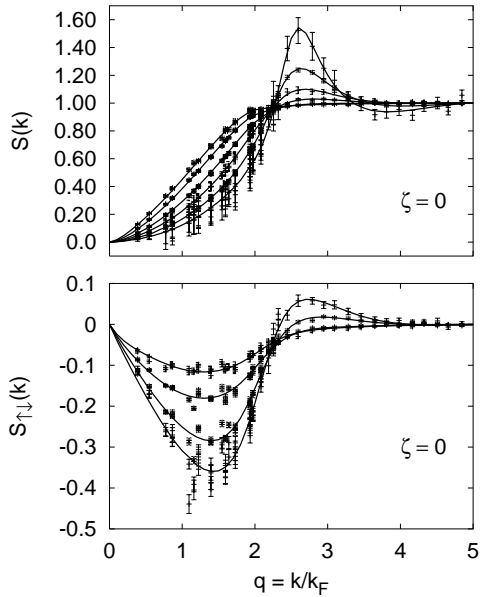} 
\caption{Upper panel: total (spin-summed) structure factor as directly obtained 
from our QMC simulations (data with error bars) and as a Fourier transform
of our analytic representation of $g^c$ (solid lines), for
$r_s\!=\!1,2,5,10,20,40$. Higher peaks correspond to larger $r_s$.
Lower panel, same comparison for the $\ud$ static structure factor:
the data with error bars are QMC simulations and the solid lines are
Fourier transforms of our analytic $g^c_{\ud}$. Here the $r_s$ values 
are 1, 2, 5, and 10 and the larger deviations from the noninteracting 
value $S_{\ud}=0$ correspond to larger $r_s$ values.}
\label{fig_Sqdmc}
\end{figure}
\begin{figure}
\includegraphics[width=6.8cm]{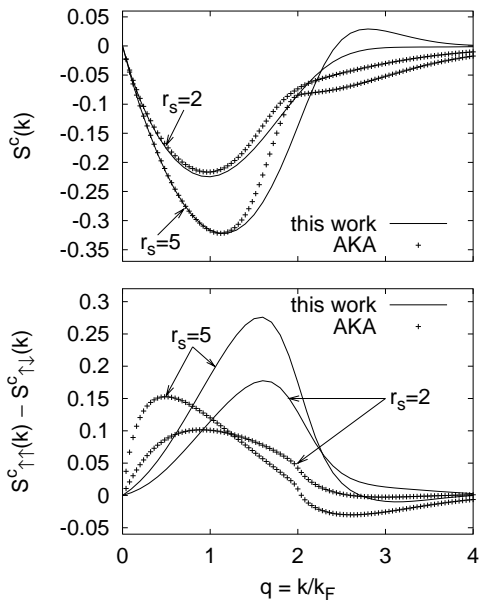} 
\caption{Upper panel: spin-summed correlation static structure factors 
from our analytic representation of $g^c$ and
from the Atwal, Khalil and Ashcroft\cite{AKA} (AKA)
dynamical local-field factor.
Lower panel: the same comparison is done for the spin channel (correlation
only, $\uu -\ud$). All curves are for $\zeta=0$.}
\label{fig_SqAKA}
\end{figure}

\section{Results}
\label{risultati}
A pictorial evidence of the quality of our analytic representation
clearly emerges from Fig.~\ref{fig_totfit}, where we show our analytic
representation for the spin-summed $g(r)$, together with the corresponding
QMC data, for $r_s=1$, 2, 5, 10, 20 and 40 and four different values
of the spin-polarization $\zeta$. 
Figure~\ref{fig_interp}, instead, shows that our analytic $g^c(r,r_s,\zeta)$
smoothly interpolates the QMC data not only as a function of $x\!=\!k_Fr$,
as e.g. shown in the previous Figure~\ref{fig_totfit}, but also as
a function of $r_s$ (upper panel) and of $\zeta$ (lower panel).
Figure~\ref{fig_uduu} summarizes similar results for $g^c_{\ud}$ at
$\zeta\!=\!0$.
The static structure factors for $\zeta\!=\!0$ are reported
in Fig.~\ref{fig_Sqdmc}. In the upper panel, we compare
the total $S(q)$ corresponding to our analytic
$g^c$ with our QMC calculation (see Sec.~\ref{sec_QMC}); 
the agreement indicates that the
long-range part ($q\to 0$ limit of $S$) of the analytic $g^c$ has been 
accurately described.
In the lower panel, we show similar results for $S_{\ud}(q)$. We see
that the long-range ($q\to 0$) spin-resolution of Eq.~(\ref{eq_Scssz0}) is
consistent with the QMC results.

Recently, Atwal, Khalil and Ashcroft\cite{AKA} (AKA) have presented 
a parametrization of the dynamical local-field factors 
(spin-symmetric, $\uu+\dd$,
and spin-antisymmetric, $\uu-\dd$) for the $\zeta\!=\!0$ 2D electron gas, 
as a function
of the wavevector $q$ and of the imaginary frequency $i\omega$.
Following the analysis carried on for the 3D electron gas by Lein,
Gross and Perdew,\cite{LGP} Asgari {\it et al.}\cite{copiaLGP}
have compared the wavector decomposition of the correlation energy
resulting from the AKA spin-symmetric local field factor 
with the one resulting from our present work, based on QMC results, 
for $r_s=1$.
In the upper panel of our Fig.~\ref{fig_SqAKA} we make a similar  
comparison (in this case at full coupling strength) for $r_s=2$ and 
$r_s=5$. In the lower panel of the same figure we also compare the results from
the AKA spin-antysimmetric local-field factors. We see that the
spin-summed AKA $S^c(q)$ is in fair agreement with our result
for $q\lesssim 1.5$, where both curves recover the exact behavior
of Eq.~(\ref{eq_Scsmallq}). The spin-antisymmetric AKA curves are, instead,
quite different from our result, even for small $q$. This discrepancy probably
comes from an inaccurate description of the high-$\omega$ behavior of the
$q\to 0$ limit of the AKA parametrization for the spin channel.\cite{quian} 
In particular,
Eq.~(26) of AKA yields a formally divergent result when combined with the
known limiting behavior\cite{raja} $S_{\ud}(q\to\infty)\propto q^{-3}$.

\section{Spin-resolved potential energy}
\label{energiapot}
\begin{figure}
\includegraphics[width=7.cm]{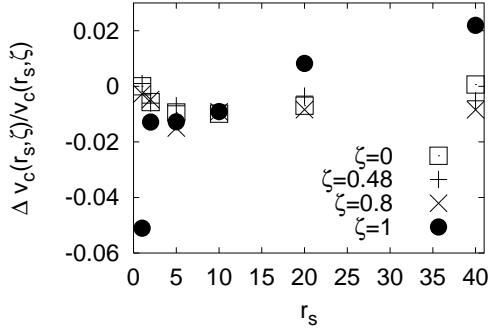}
\caption{Relative error $\Delta v_c(r_s,\zeta)/v_c(r_s,\zeta)=
(v_c-v_c^{\rm INT})/v_c$
between $v_c$ calculated using the r.h.s.~of Eq.~(\ref{eq_vc}) (with
$\ec$ from Ref.~\onlinecite{AMGB}), and $v_c^{\rm INT}$, obtained
by numerical integration of $g^c$ (see text).}
\label{fig_errvc}
\end{figure}

The correlation part of the potential energy, $v_c(r_s,\zeta)$ of Eq.~(\ref{eq_vc}),
can be divided into $\uu$, $\dd$, and $\ud$ contributions, such that
$v_c=v_c^{\uu}+v_c^{\dd}+v_c^{\ud}$. These spin-resolved components of 
$v_c$ are important ingredients for
the study and construction of dynamical exchange-correlation 
potentials in the spin channel.\cite{vignale,quian}
They can be written as
the expectation value of the Coulomb potential $1/r$  
on the spin-resolved $g_{\sigma\sigma'}^c$,
\beq
v_c^{\sigma\sigma'}(r_s,\zeta)=
\frac{(2-\delta_{\sigma\sigma'})}{\sqrt{2}\,r_s}\frac{n_\sigma n_{\sigma'}}{n^2}
\int_0^{\infty} g^c_{\sigma\sigma'}(x,r_s,\zeta)\, dx.
\label{eq_vcss}
\eeq 
We have evaluated 
the r.h.s.~of Eq.~(\ref{eq_vcss}) by numerical integration 
of our QMC data for $g_{\sigma\sigma'}^c(x,r_s,\zeta)$, at $\zeta=0,0.48,
0.8$, and $r_s=1,2,5,10,20,40$. This means that the integration in the
r.h.s.~of
Eq.~(\ref{eq_vcss}) has been truncated at $L/2$, where $L$ is the
side of the simulation cell (in our case $L/2\sim 6\,r_s$).
The resulting $v_c^{\sigma\sigma'}$ are thus
affected by the finite-size error, since they correspond 
to systems with fixed number of particles
(see Sec.~\ref{sec_QMC}) and an infinite-size extrapolation
is not available in this case.
One can get an idea of the magnitude of such error
by using the same numerical-integration procedure for the spin-summed
$g^c$, and then comparing the results with the corresponding
thermodynamic limit,
the last term of Eq.~(\ref{eq_vc}), combined with our\cite{AMGB} 
$\ec(r_s,\zeta)$. The relative error between
the two evaluations of $v_c$ is reported in Fig.~\ref{fig_errvc}:
it is of the order of few percents. 
At $\zeta =1$, Figure~\ref{fig_errvc} disproofs Eq.~(10) of 
Ref.~\onlinecite{MM}, which predicts a different
virial theorem for the fully polarized system.
\begin{figure}
\includegraphics[width=7.cm]{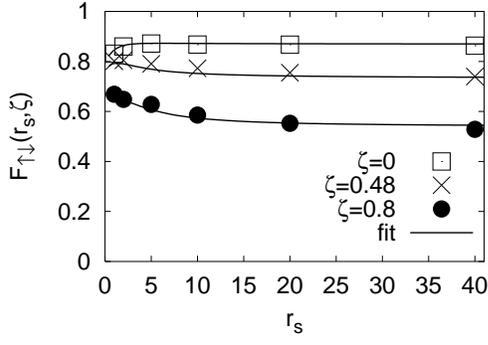}
\caption{Fraction of $\ud$ contribution to the correlation part of
the potential energy, $F_{\ud}=v_c^{\ud}/v_c$.}
\label{fig_Fud}
\end{figure}
 
We have parametrized our spin-resolved $v_c^{\sigma\sigma'}(r_s,\zeta)$ 
as\cite{GP3}
\beq
v_c^{\sigma\sigma'}(r_s,\zeta)=F_{\sigma\sigma'}(r_s,\zeta)\, v_c(r_s,
\zeta).
\eeq
The fractions $F_{\sigma\sigma}(r_s,\zeta)$ for parallel spins 
are well represented by
\begin{eqnarray}
F_{\uu}(r_s,\zeta)& = & F_{\uu}^{\rm HD}(\zeta)+\left[w_1(\zeta)\,r_s+
w_2(\zeta)\, r_s^2\right] \times \nonumber \\
& & \times \log\left(1+\frac{w_3(\zeta)}{r_s^2}\right),
\label{fit_Fuu}
\end{eqnarray}
where the high-density $F_{\uu}^{\rm HD}$ is given by Seidl,\cite{mike} 
\begin{eqnarray}
F_{\uu}^{\rm HD}(\zeta)& = & \frac{-19.54\,(1+\zeta)}
{153.38 {\cal{F}}(\zeta)-192.46}, \\
{\cal{F}}(\zeta) & = & \frac{(1+\zeta)\log(1+\zeta)+(1-\zeta)\log(1-\zeta)}
{2\log(2)} +\nonumber \\
& & + 0.0636\,\zeta^2-0.1024\,\zeta^4+0.0389\,\zeta^6,
\end{eqnarray}
and the functions $w_i(\zeta)$ have been obtained by fitting our data for
$v_c^{\uu}(r_s,\zeta)$ for $\zeta=-0.8,-0.48,0.,0.48,0.8$ (the negative
$\zeta$ values corresponding to the $\dd$ data),
\begin{eqnarray}
w_1(\zeta) & = & (1-\zeta)\,(-0.006-0.03\,\zeta) \\
w_2(\zeta) & = & (1-\zeta)\,(-0.01+0.03\,\zeta) \\
w_3(\zeta) & = & 3.6\,(1+\zeta)^4.
\label{eq_iwi}
\end{eqnarray}
Equations~(\ref{fit_Fuu})-(\ref{eq_iwi}) completely determine
the spin resolution of $v_c(r_s,\zeta)$, since $F_{\dd}(r_s,\zeta)=
F_{\uu}(r_s,-\zeta)$ and $F_{\ud}=1-F_{\uu}-F_{\dd}$.

In Fig.~\ref{fig_Fud} we show our numerical results for the antiparallel-spin fraction 
$F_{\ud}(r_s,\zeta)$, together with our fitting function; the relative errors on the fit 
of $v_c^{\ud}$ (not shown) are of the
same order of magnitude of those of Fig.~\ref{fig_errvc}. We see that the correlation part of
the potential energy is completely dominated by the $\ud$ contribution, even for $\zeta$
as high as 0.8.

\section*{Acknowledgments}
We thank R. Asgari, S. De Palo, Z. Qian, S. V. Kravchenko, C. Tanguy,
G. Vignale, and  W. Yang for 
useful discussions, and gratefully acknowledge financial support from the Italian
Ministry of Education, University and Research (MIUR) through COFIN 2003-2004
and the allocation of computer resources from INFM Iniziativa Calcolo Parallelo.

\appendix
\section{Long-Range scaling}
\label{app_LR}
In this appendix we describe the construction of the long-range part
of our $g^c$. We follow the same procedure used
for the 3D case in Ref.~\onlinecite{WP}, where the interested reader
can find more details and comments on the relevant physics.
Here we briefly recall the main equations and emphasize the 
differences between the 3D and the 2D case.
As for the 3D case,\cite{GGA} the results of this appendix can be used 
in the construction
of a generalized-gradient approximation for a 2D correlation energy functional.
 
Following Ref.~\onlinecite{WP}, we seek a scaling law
for the long-range part of $g^c$. We call ``long-range'' part
the oscillation-averaged asymptotic behavior of $g^c$ for large $x=\kf r$, or
equivalently, the behavior of $S^c$ for small $q=k/\kf$.
From Eq.~(\ref{eq_Scsmallq}), we see that such a scaling law has the form
\beq
S^c(q\to 0,r_s,\zeta) \to \sqrt{2}\, r_s\, \phi^3\, f(z,\zeta),
\label{eq_scaling}
\eeq
where
\beq
z=\frac{k}{k_{TF} \phi^2} = \frac{q}{\phi^2\sqrt{2}\,{r_s}},
\eeq
is a variable on the scale
of the Thomas-Fermi wavevector, $k_{TF}$ (which does not depend
on $r_s$ in the 2D case), and the function $f(z,\zeta)$ has 
the small-$z$ expansion (independent of $\zeta$)
\beq
f(z\to 0,\zeta) = -\tfrac{2}{\pi}z+\tfrac{1}{\sqrt{2}}z^{3/2}+O(z^2).
\label{eq_fz}
\eeq
The random-phase approximation (RPA) exactly recovers 
Eqs.~(\ref{eq_scaling})-(\ref{eq_fz}).
As in the 3D case,\cite{WP} we can thus obtain the function
$f(z,\zeta)$ from RPA. Its (wrong) short-range behavior will be
 quenched in our parametrization of $g^c$ by the cutoff
function $\fcut(x)$ of Eq.~(\ref{eq_fcut}).
\begin{figure}
\includegraphics[width=7cm]{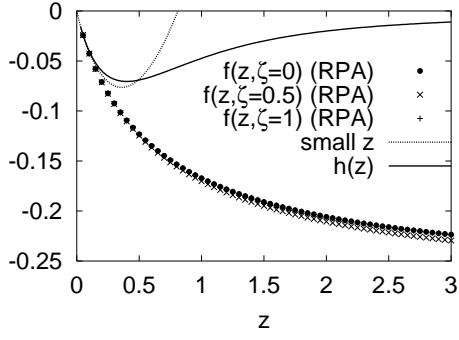} 
\caption{The function $f(z,\zeta)$ [see Eq.~(\ref{eq_scaling})], 
evaluated within RPA, 
for three different values of the
spin polarization parameter $\zeta$. Also shown: the exact small-$z$
behavior of Eq.~(\ref{eq_fz}) and the function $h(z)$ of
Eq.~(\ref{eq_hz}).}
\label{fig_fz}
\end{figure}

We thus evaluated the function $f(z,\zeta)$ via the standard
RPA equation
\beq
S^c_{\rm RPA}(q,r_s,\zeta) = -\int_0^{\infty} d\omega
\frac{(\beta_\u+\beta_\d)^2}{q\kf/\pi-(\beta_\u+\beta_\d)},
\label{eq_ScRPA}
\eeq
with 
\beq
\beta(q,\omega)=-\frac{2}{\pi q}\left[\frac{q}{2}-
{\rm Re}\left(\sqrt{\left(\frac{q}{2}+i\frac{\omega}{q}\right)^2-1}\right)\right],
\eeq
and
\beq
\beta_{\u}=\beta\left(\frac{q}{\sqrt{1+\zeta}},\frac{\omega}{1+\zeta}
\right), \qquad
\beta_{\d}=\beta\left(\frac{q}{\sqrt{1-\zeta}},\frac{\omega}{1-\zeta}
\right).
\eeq
The resulting $f(z,\zeta)$ has the small-$z$ expansion
of Eq.~(\ref{eq_fz}), and for large $z$ is equal to
\beq
f(z\to \infty,\zeta) = A(\zeta)+\frac{B(\zeta)}{z}+...
\label{eq_largez}
\eeq
Equation~(\ref{eq_largez}) corresponds, in real space, to a divergent
short-range
behavior. As in three dimensions,\cite{WP} with a suitable cutoff
Eq.~(\ref{eq_largez}) recovers the RPA high-density expansion
of the correlation energy, $\ec^{\rm RPA}(r_s\to 0,\zeta) = 
a_{\rm RPA}(\zeta)+b_{\rm RPA}(\zeta)r_s\log r_s+O(r_s)$. However, unlike
the 3D case,
in two dimensions the RPA correlation energy
is not exact even in the $r_s\to 0$ limit.\cite{raja}  
Thus, while in  three dimensions\cite{WP,PW,GP2}
the large-$z$ behavior of $f(z,\zeta)$ was important to recover
the exact high-density limit of $\ec$, there is no need in 2D
to keep Eq.~(\ref{eq_largez}) in our parametrization. Moreover,
as in three dimensions,\cite{WP} we find that the $\zeta$ dependence of
$f(z,\zeta)$ is very weak (see Fig.~\ref{fig_fz}), so that 
we can replace $f(z,\zeta)$ with
$f(z,0)$ [this is exact in the ``important'' part of $f$, i.e., the 
small-$z$ regime of Eq.~(\ref{eq_fz})]. We thus define the function $h(z)$ 
\beq
h(z)=f(z,0)-A(0)-\frac{B(0)}{\sqrt{z^2+[B(0)/A(0)]^2}}
\label{eq_hz}
\eeq
where $A(0)=-0.272076$ and $B(0)=10/\pi-3$ correspond to the large-$z$ 
expansion for $\zeta=0$ of Eq.~(\ref{eq_largez}). 
As shown in Fig.~\ref{fig_fz}, the function $h(z)$ 
has the same small-$z$ behavior of $f(z,\zeta)$, but goes to zero
when $z\to\infty$, which corresponds to  
a less diverging short-range part in real space. 
The Fourier transform
of $h(z)$ defines the function $f_1(v)$ of Eq.~(\ref{eq_glr}),
\beq
\frac{f_1(v)}{v}=\int_0^{\infty} h(z)\,z\,J_0(vz)\,dz,
\eeq
where $v=\sqrt{2} r_s \phi^2 x$ is the appropriate real-space 
scaled variable. The function $f_1(v)$ has been evaluated
numerically, and then parametrized as
\beq
f_1(v)=\frac{b_1\,v^{1/2}+b_2\,v+b_3\,v^{3/2}+b_4\, v^2+b_5\,v^{5/2}
+b_6\,v^3}{(v^2+b_0^2)^{5/2}}
\label{eq_f1v}
\eeq
with
\bear
b_6 & = & 2/\pi \label{eq_b6}\\
b_5 & = & -\frac{9}{4\pi}\frac{1}{\sqrt{2}}\left[\Gamma\left(\frac{3}{4}\right)
\right]^2 \\
b_4 & = & -3\,b_0\Biggl[\frac{b_1}{2b_0^{5/2}}{\rm B}\left
(\frac{3}{4},\frac{7}{4}\right)
+\frac{b_2}{3b_0^2}+\frac{b_3}{2b_0^{3/2}}
{\rm B}\left(\frac{5}{4},\frac{5}{4}\right)\nonumber \\
& & 
+\frac{b_5}{2b_0^{1/2}}{\rm B}\left(\frac{3}{4},\frac{7}{4}\right)
+\frac{2}{3}b_6\Biggr],
\label{eq_b4}\\
\vphantom{\biggl[} b_3& = &-22 \quad\quad b_2=61 \quad\quad b_1=-64 
\quad\quad b_0=3.46 \, .
\nonumber
\eear
Eqs.~(\ref{eq_b6})-(\ref{eq_b4}) guarantee that the Fourier
transform of $f_1(v)/v$ satisfies Eq.~(\ref{eq_fz}). $\Gamma(x)$
and ${\rm B}(x,y)$ are the standard Gamma and Beta functions:\cite{abramovitz}
$\Gamma(3/4)\approx 1.225416702$, ${\rm B}(3/4,7/4)\approx 0.8472130848$,
${\rm B}(5/4,5/4)\approx 0.6180248924$.

\section{Spin-resolution of the long-range part ($\zeta=0$)}
\label{app_LRss}
The long-range part of $g^c_{\sigma\sigma'}(x,r_s,\zeta=0)$
has been simply approximated with Eq.~(\ref{eq_gsslr}), where
the function $\tilde{f}_1(v,\alpha)$ is obtained from $f_1(v)$
of Eq.~(\ref{eq_f1v})
by replacing $b_6$ with $\tilde{b}_6=2(1/\pi+\alpha)$,
and consequently changing  $b_4$ according to Eq.~(\ref{eq_b4}). 
In this way, the corresponding $S^c_{\sigma\sigma'}(q,r_s,\zeta=0)$
exactly recovers Eq.~(\ref{eq_Scssz0}).


\begin{thebibliography}{100}
\bibitem{nota} For the low-density limit, we have simply fitted the frequency and
phase of our oscillating term to a model pair-distribution function built from
gaussian orbitals centered on the sites of a triangular lattice. In the present
work we are dealing with the fluid state, not with the crystal, and the {\it damping}
of the oscillations is completely different in the two cases. But the {\it position} of
the maxima, which is practically the same in the fluid and crystalline state, is the
only information which we extract from this model.
\end{thebibliography}
\end{document}